# A Cluster-Based Computational Thermodynamics Framework with Intrinsic Chemical Short-Range Order: Part I. Configurational Contribution

Chu-Liang Fu, Rajendra Prasad Gorrey, Bi-Cheng Zhou[*]

Department of Materials Science and Engineering, University of Virginia, Charlottesville, Virginia 22904, USA

**Abstract**

Exploiting Chemical Short-Range Order (CSRO) is a promising avenue for manipulating the properties of alloys. However, existing modeling frameworks are not sufficient to predict CSRO in multicomponent alloys (>3 components) in an efficient and reliable manner. In this work, we developed a hybrid computational thermodynamics framework by combining unique advantages from Cluster Variation Method (CVM) and CALculation of PHAse Diagram (CALPHAD) method. The key is to decompose the cumbersome cluster variables in CVM into fewer site variables of the basic cluster using the Fowler-Yang-Li (FYL) transform, which considerably reduces the number of variables that must be minimized for multicomponent systems. CSRO is incorporated into CALPHAD with a novel cluster-based solution model called FYL-CVM. This new framework brings more physics into CALPHAD while maintaining its practicality and achieves a good balance between accuracy and computational cost. It leverages statistical mechanics to yield a more physical description of configurational entropy and opens the door to cluster-based CALPHAD database development. The application of the FYL-CVM model in a prototype fcc AB alloy demonstrates its capability to correctly reproduce the essential features of the phase diagram and thermodynamic properties. The hybrid CVM-CALPHAD framework represents a new methodology for thermodynamic modeling that enables atomic-scale order to be exploited for materials design.

**Keywords:** CALPHAD; Chemical short-range order; Thermodynamics; Cluster Variation Method

---

[*] Corresponding author: Bi-Cheng Zhou (bicheng.zhou@virginia.edu)

# 1. Introduction

Recent years have witnessed enormous growth of the field of high entropy alloys or complex concentrated alloys (CCAs) [1,2]. CCAs bring immense opportunities for property tuning due to their vast composition space. They have long been thought to mostly contain a single-phase random solid solution but recent studies have found that order-disorder transitions [3,4] and Chemical Short-Range Order (CSRO) [5–8] are common in CCAs. The existence of CSRO in CCAs was directly verified through transmission electron microscopy [9,10]. CSRO is also reported to affect the stacking-fault energy and dislocation mobility [11,12] as a major factor in controlling mechanical properties in CCAs. The ubiquitous atomic-scale order is being realized by the materials community as a new knob for property manipulation of materials [13–15]. Therefore, exciting opportunities exist in exploiting CSRO for property enhancement in CCAs and other alloys. However, tools for reliably and efficiently predicting CSRO in multicomponent alloys (>3 components) do not exist.

Known for its simplicity and efficiency, the CALPHAD (CALculation of PHAse Diagram) method is a leading method for thermodynamic modeling and calculations of phase equilibria in multicomponent materials [16,17]. Instead of relying solely on first-principles inputs, experimental thermochemical and phase equilibria data are often combined with theoretical data to parameterize a self-consistent thermodynamic model in CALPHAD modeling. Thermodynamic databases of realistic multicomponent alloys can thus be established. The importance of these databases lies in not only their capability to predict phase equilibria and plot complex phase diagrams, but also their role as inputs for kinetic models such as the phase-field model [18]. CALPHAD has already been used to understand phase stability in CCAs [19–22]. Closed-form free energy models are at the core of CALPHAD, with which phase diagrams can be efficiently calculated without statistical sampling. However, the prevailing solution model used in CALPHAD, the sublattice model or compound energy formalism [23], is an empirical mean-field model based on the Bragg-Williams (ideal entropy of mixing) approximation. This makes CALPHAD inadequate for properly describing order-disorder transformations or CSRO in alloys, such as the Guinier-Preston zones or nanoscale clusters [24], which are critical for alloy mechanical properties. Therefore, it is highly desirable to develop a solution model with intrinsic CSRO for CALPHAD.

Throughout the history of alloy theory, many solution models have been developed to describe CSRO. In the 1940s, Yang [25,26] and Li [27,28] derived a generalized quasi-chemical model of order in superlattices (the Yang-Li approximation). Pioneered by Kikuchi [29] in 1951, Cluster Variation Method (CVM) [30–32], a direct approximation method to the free energy describes the ordering structures by a hierarchical cluster formalism. However, the minimization of free energy in the CVM framework causes a huge computational burden, especially for multicomponent alloys. In recent years, cluster expansion coupled with the Monte Carlo (MC) method [33,34] has been widely adopted to calculate phase diagrams for ordering alloys. However, this method cannot be combined with experimental data to produce quantitatively accurate phase diagrams for technological applications because it is limited by the energetic accuracy of first-principles inputs [35]. CVM has ultimately decreased in popularity in favor of cluster expansion using correlation functions based on explicit formulation of linear equations relating different cluster configurations, because the intricacy of constraints in CVM renders it cumbersome to formulate for large clusters [36]. Furthermore, the combinatorial explosion of configuration variables tends to limit CVM and cluster expansion to binary or ternary systems. To address this issue, Oates later revived the Yang-Li approximation [25–28] by proposing the Cluster/Site Approximation (CSA) [37,38]. The key advantage of CSA compared to CVM is that the number of minimizing variables to approach equilibrium is dramatically reduced. This feature makes CSA a candidate of the solution model suitable for CALPHAD. However, the major assumption in CSA that the basic clusters must be of non-interference leads to significant inaccuracy [37]. To improve its accuracy, Oates inserted a modeling parameter without clear physical meaning into the CSA entropy formula to mimic the contribution of the cluster interactions [38]. The quasi-chemical model [39] and the subsequent modified version [40], which are developed by Pelton and co-workers and commonly used in CALPHAD, consider the pair interactions but ignore other types of short-range interactions. Most recently, van de Walle et al. [41] proposed a CSRO correction in the context of high-throughput CALPHAD using truncated polynomial expansions of the CVM pairwise entropy, in order to make full use of the high-throughput results of special quasi-random structures [42]. However, it is still based on the sublattice model with incomplete CSRO.

It must be emphasized that CSRO plays a significant role in relative phase equilibrium although its contribution to the alloy's total free energy difference is usually small compared with differences caused by the crystal structure change [43]. Equilibrium between two different crystal structures usually involve phases that are far from the ordering transition on their respective sublattices. Therefore, CSRO effects are often incorporated into the sublattice models by means of a phenomenological expansion of the excess free energy with reciprocal interaction parameters [44] by CALPHAD practitioners, but this empirical approach to the description of configurational free energy has serious limitations. The most significant one is that the configurational entropy cannot be properly approximated by a polynomial expansion over extended temperature and composition ranges [43]. On the contrary, using a cluster-based solution model connects CALPHAD with the mainstream statistical mechanics [45] and naturally accounts for CSRO within clusters. Moreover, if a statistical-mechanics-based model is adopted to describe the configurational contribution to free energy, the non-configurational contributions (elastic, vibrational, electronic) also need to be considered [31].

To address the shortcomings of existing models, in this *Part I* paper, we present a novel cluster-based configurational thermodynamic model that enables the prediction of CSRO and fulfils all the requirements of a thermodynamic solution model for CALPHAD: scalable to multicomponent systems, accurate, low computational cost, and easy-to-use. The non-configuration contributions (elastic, vibrational, and electronic) will be discussed in a *Part II* paper. This paper is organized as follows. In Section 2, we present a theoretical derivation of the proposed novel cluster-based solution model. The predicted prototype phase diagrams and thermodynamic properties based on different models are presented and compared in Section 3. Discussion on the features of the model is provided in Section 4, with conclusion and outlook in Section 5.

## 2. The FYL-CVM model

The key to our novel solution model is a mathematical transform called the Fowler-Yang-Li (FYL) transform, which was coined by Oates [46] to honor Fowler [47], Yang [25,26], and Li [27,28] who pioneered this idea. The application of the FYL transform to CVM was hinted at one of Oates' papers [46] but was never elaborately derived nor tested. In this work, we propose a new model by applying FYL transform to the cluster variables of CVM keeping the original name of FYL-CVM [46] for our model. The number of minimizing variables in CVM is significantly reduced after the FYL transform, making it feasible to extend to multicomponent alloys. In this section, we will demonstrate the steps to derive the FYL-CVM model.

## 2.1. Model setup

The Gibbs free energy of a phase in an alloy containing $n$ components is expressed as:

$$G = \sum_{i=1}^{n} N x_i G_i^\circ + \Delta G_{mix}^{conf} + \Delta G_{mix}^{nconf} \tag{1}$$

where $N$ is the total number of atoms, $G_i^\circ$ is the molar Gibbs free energy of pure component $i$ (i.e., lattice stability) as a reference state, which can be taken from the SGTE database of pure elements [48]. $\Delta G_{mix}^{conf}$ is the configurational free energy of mixing, and $\Delta G_{mix}^{nconf}$ is the non-configurational free energy of mixing including elastic, vibrational, and electronic contributions. Herein, we take a hypothetical binary AB alloy on an fcc lattice as a model system to demonstrate the derivation of our novel cluster-based solution model. This model system uses the 1$^{st}$ nearest-neighbor tetrahedron as the basic cluster (i.e., maximum cluster), as shown in **Figure 1**. The four atomic sites $s$ in the tetrahedron basic cluster are labeled with $s$=1, 2, 3, and 4. The list of the symbols and notations used throughout the paper is summarized in the Appendix.

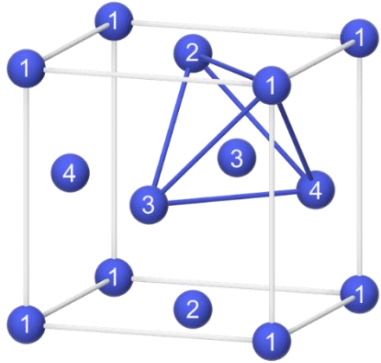

**Figure 1.** The tetrahedron basic cluster with four sites 1~4 on an fcc lattice

## 2.2. Model derivation

The Gibbs free energy $G$ is equal to the Helmholtz free energy $F$ in this work because we do not consider the external pressure contribution to free energy and there is no volume change. To calculate the temperature-composition phase diagrams, the Helmholtz free energy of mixing $\Delta F_{mix}$ needs to be minimized with fixed compositions in the canonical (NVT) ensemble. However, the grand potential $\Omega$ is often adopted as the minimizing variable while fixing the chemical potentials $\mu$ rather than the compositions $x$ in practical CVM free energy minimization [49]. The Lagrange multiplier method accounting for the normalization constraint of cluster probability variables is usually used [50]. Based on the statistical mechanics definition of the grand canonical ensemble (μVT), we have

$$\Delta G_{mix}^{conf} = \Delta F_{mix}(N, V, T) = \Delta \Omega_{mix}(\mu, V, T) + \Delta \mu_{mix} N \tag{2}$$

where $T$ is the temperature, $\Delta\Omega_{mix}$ is the grand potential of mixing to be minimized, and $\Delta\mu_{mix}$ is the change of chemical potential (i.e., molar Gibbs free energy) upon mixing. The original derivation by Yang [25,26] using the FYL transform is based on canonical ensemble. A more succinct version of Yang's derivation is presented in Supplementary Materials. Our derivation instead utilizes grand canonical ensemble, which reveals the essence of the FYL transform in a clearer way and avoids the complex Legendre transformation used in Yang's original derivation [25,26].

Next, we apply the CVM hierarchical formalism [51,52] to perform a cumulant expansion of the total free energy as a weighted summation of contributions from basic clusters and their subclusters. Such a CVM hierarchy, which evolved over many years into its current form [51,53–55], can avoid overcounting during the statistical calculation of the clusters [56]. In lieu of separating F into energy and entropy terms as in previous CVM formulations [50,57], in this work, our derivation directly expand the $\Delta\Omega_{mix}$ and $\Delta\mu_{mix}$ in a CVM hierarchical formalism [51,52]. The purpose is to expose the cluster chemical potential terms which we will apply the FYL transform to. Based on the reason above, we can expand the $\Delta\Omega_{mix}$ and $\Delta\mu_{mix}$ in the CVM hierarchical formalism [58] as

$$\Delta F_{mix} = \Delta\Omega_{mix} + \Delta\mu_{mix} N \stackrel{CVM}{\Longrightarrow} N \sum_{\alpha \in R} a_\alpha m_\alpha (\Omega_\alpha + \mu_\alpha) \tag{3}$$

The sum is over the set $R$ of all symmetry-equivalent clusters $\alpha$ (also called "orbits"). $\Omega_\alpha$ is the grand potential for cluster type $\alpha$. $\mu_\alpha$ is the overall chemical potential for cluster type $\alpha$. $a_\alpha$ and $m_\alpha$ are coefficients to eliminate overcounting in various clusters' contributions. The mathematical foundation of $a_\alpha$ and $m_\alpha$ in CVM is the exclusion-inclusion principle [59] with the Möbius inversion [60,61]. $a_\alpha$ is the Möbius numbers and can be ±1 [60,61]. The multiplicity $m_\alpha$ is the number of symmetry-equivalent clusters of type $\alpha$ per lattice site, which solely depends on the geometry of the lattice [62]. $\gamma_\alpha = a_\alpha m_\alpha$ is often referred as the Kikuchi-Barker coefficient [62]. This CVM expansion defines a variational functional to reach equilibrium through minimization.

It is more convenient to minimize the cluster grand potential $\Omega_\alpha$ by varying $\mu_\alpha$ of different configurations than varying composition [63]. For the cluster grand potential $\Omega_\alpha$, we have

$$\Omega_\alpha = -k_B T \ln z_\alpha \tag{4}$$

where $z_\alpha$ is the cluster grand partition function. The overall chemical potential for cluster type $\alpha$ can be expressed as

$$\mu_\alpha = \sum_c \rho_c \mu_\alpha(c) \tag{5}$$

where $c$ denotes all the distinctive atomic configurations in cluster $\alpha$ and $\rho_c$ is the corresponding cluster probability.

Based on the properties of the grand canonical ensemble, the cluster grand partition function can be written as

$$z_\alpha = \sum_c \exp\left(\frac{\mu_\alpha(c) - \varepsilon_c}{k_B T}\right) \tag{6}$$

where $k_B$ is the Boltzmann constant. The cluster probabilities are calculated by

$$\rho_c = \frac{1}{z_\alpha} \exp\left(\frac{\mu_\alpha(c) - \varepsilon_c}{k_B T}\right) \tag{7}$$

In CVM, the free energy functional must be minimized through varying these cluster probabilities $\rho_c$ to find the equilibrium state. The CSRO can be reflected by the calculated $\rho_c$. $\varepsilon_c$ is the energy of a basic cluster which is a modeling parameter. $\mu_\alpha(c)$ is the variable used in free energy minimization.

All the above equations are generally applicable to any lattice. With regard to the binary AB alloy on an fcc lattice with tetrahedron as the basic cluster, the cluster grand partition function becomes

$$z_\alpha = \sum_{i,j,k,l} \exp\left(\frac{\mu_\alpha(ijkl) - \varepsilon_{ijkl}}{k_B T}\right) \tag{8}$$

where $ijkl$ is the notation to represent the complete enumerated set of cluster configurations consisting of atomic species A or B occupying the indicated sites 1~4 in the basic tetrahedron cluster. The set includes: AAAA, AAAB, AABA, ABAA, BAAA, BBAA, ABBA, AABB,

BABA, ABAB, BAAB, BBBA, ABBB, BBAB, BABB, and BBBB. The cluster probabilities are calculated by

$$\rho_{ijkl} = \frac{1}{Z_\alpha} \exp\left(\frac{\mu_\alpha(ijkl) - \varepsilon_{ijkl}}{k_B T}\right) \tag{9}$$

The use of grand canonical ensemble reveals the essence of the CVM, that it a variational technique that consists in minimizing a free energy functional where the chemical potentials (activities) of the basic clusters are varied to minimize $\Delta\Omega_{mix}$. All the derived formulas in Section 2.2 should in principle work for conventional CVM as well. The key difference between CVM and FYL-CVM is the application of the so-called FYL transform.

2.3. *The FYL transform*

The FYL transform is an ansatz to reduce the number of minimizing variables in the CVM equilibrium calculations and is applied directly on the basic cluster chemical potential $\mu_\alpha(c)$ (not sub-clusters). Based on Fowler's theory of gaseous atom/molecule equilibrium [47], if the equilibrium constant and the atom/molecule mass balance are known, the Helmholtz energy can be expressed in terms of the atomic concentrations rather than the molecular concentrations. This idea was later applied to clusters in an fcc crystal by Yang [25,26] and Li [27,28]. They derived a generalized quasi-chemical model of order in superlattices by equilibrating the cluster distribution with a fictitious atomic gas. Oates revived the Yang-Li approximation as CSA [37] and was the first to call this ansatz as "FYL transform" [64]. For the binary AB alloy on an fcc lattice with tetrahedron as the basic cluster, the idea of FYL transform and its analogy to gaseous chemical equilibria is illustrated in **Figure 2**.

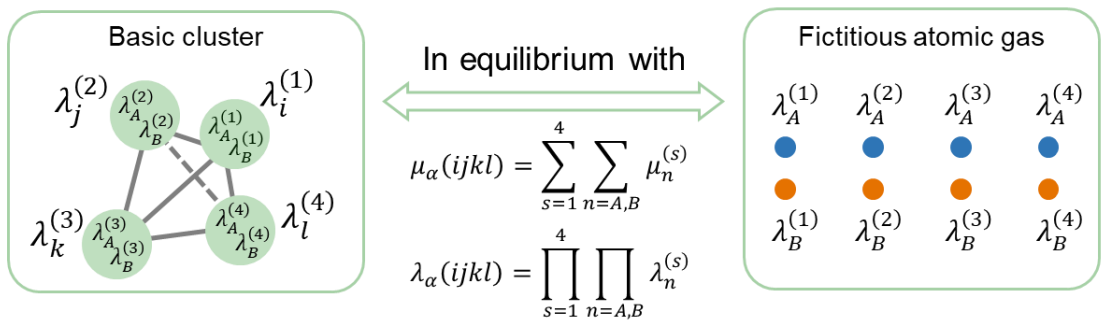

**Figure 2.** Illustration of the idea of FYL transform and its analogy to gaseous chemical equilibria.

The essence of FYL transform is to decompose the basic cluster chemical potential in the cluster grand partition function into the summation of individual chemical potentials of the component on the site of the basic cluster. The former depends on the cluster probabilities while the latter on the site variables. After the FYL transform, the basic cluster chemical potential $\mu_\alpha(c)$ is expressed as

$$\mu_\alpha(c) \xrightarrow{FYL\ transform} \sum_{s,n} \mu_n^{(s)} \tag{10}$$

where $\mu_n^{(s)}$ is the site chemical potential of component $n$ on site $s$. We also have

$$\mu_n^{(s)} = k_B T \ln \lambda_n^{(s)} \tag{11}$$

where the site variable $\lambda_n^{(s)}$ is the absolute activity of component $n$ on site $s$ [65]. Combining Eqs. (10) and (11)(11)(12), we have

$$\mu_\alpha(c) = \sum_{s,n} k_B T \ln \lambda_n^{(s)} \tag{12}$$

If we define $\lambda_\alpha(c)$ as an effective absolute activity of the basic cluster, the FYL transform equivalently be expressed as

$$\lambda_\alpha(c) \xrightarrow{FYL\ transform} \prod_{s,n} \lambda_n^{(s)} \tag{13}$$

Substituting Eqs.(12) and (13) into Eqs.(8) and (9), the cluster grand partition function $z_\alpha$ and the corresponding basic cluster probability $\rho_{ijkl}$ can be expressed as

$$z_\alpha = \sum_c \left[ \prod_{s,n} \lambda_n^{(s)} \exp\left(-\frac{\varepsilon_c}{k_B T}\right) \right] \tag{14}$$

$$\rho_c = \frac{1}{z_\alpha} \prod_{s,n} \lambda_n^{(s)} \exp\left(-\frac{\varepsilon_c}{k_B T}\right) \tag{15}$$

In the case of fcc AB alloy with a tetrahedron basic cluster, the FYL-transformed basic cluster $\mu_\alpha(ijkl)$ can be expressed as

$$\mu_\alpha(ijkl) = \sum_{s=1}^{4} \sum_{n=A,B} \mu_n^{(s)} = \mu_A^{(1)} + \mu_B^{(1)} + \mu_A^{(2)} + \mu_B^{(2)} + \mu_A^{(3)} + \mu_B^{(3)} + \mu_A^{(4)} + \mu_B^{(4)} \tag{16}$$

The above equation can be further simplified by eliminating all the $\mu_B^{(s)}$ because only the relative value of $\mu_n^{(s)}$ is meaningful in the variation process to reach equilibrium [63]. Setting B as a reference, which leads to $\mu_B^{(s)} = 0$ ($\lambda_B^{(s)} = 1$) for all $s$, we have

$$\begin{aligned}\mu_\alpha(ijkl) &= \mu_i^{(1)} + \mu_j^{(2)} + \mu_k^{(3)} + \mu_l^{(4)} \\ &= k_B T\left(\ln \lambda_i^{(1)} + \ln \lambda_j^{(2)} + \ln \lambda_k^{(3)} + \ln \lambda_l^{(4)}\right) \\ &= k_B T \ln\left(\lambda_i^{(1)} \lambda_j^{(2)} \lambda_k^{(3)} \lambda_l^{(4)}\right)\end{aligned} \quad (17)$$

If $i,j,k,l = B$, $\lambda_B^{(s)} = 1$. If $i,j,k,l = A$, $\lambda_A^{(s)}$ survives. For example, $\mu_\alpha(AAAB) = k_B T \ln\left(\lambda_A^{(1)} \lambda_A^{(2)} \lambda_A^{(3)}\right)$. The basic cluster grand partition function $z_\alpha$ and the corresponding cluster probability $\rho_{ijkl}$ can thus be expressed as

$$z_\alpha = \sum_{i,j,k,l} \lambda_i^{(1)} \lambda_j^{(2)} \lambda_k^{(3)} \lambda_l^{(4)} \exp\left(-\frac{\varepsilon_{ijkl}}{k_B T}\right) \quad (18)$$

$$\rho_{ijkl} = \frac{1}{z_\alpha} \lambda_i^{(1)} \lambda_j^{(2)} \lambda_k^{(3)} \lambda_l^{(4)} \exp\left(-\frac{\varepsilon_{ijkl}}{k_B T}\right) \quad (19)$$

With the FYL transform, the configuration variables in cluster calculations of free energy are changed from cluster probabilities $\rho_{ijkl}$ to the considerably fewer site probabilities $\lambda^{(s)}$. The minimizing variables are reduced from the order of $n^s$ in CVM to the order of $n \times s$ in FYL-CVM, thus resolving the infamous "combinatorial explosion" problem in CVM when dealing with multicomponent system with large basic clusters. This paves the way for FYL-CVM to be compatible with CALPHAD in describing multicomponent systems.

The equilibrium site fractions $x_A^{(s)}$ for A on site $s$ must satisfy the CVM superposition relations [66], which are expressed as a sum of the basic cluster probabilities on the corresponding site:

$$\begin{aligned}x_A^{(1)} &= \sum_{j,k,l} \rho_{Ajkl} \\ x_A^{(2)} &= \sum_{i,k,l} \rho_{iAkl} \\ x_A^{(3)} &= \sum_{i,j,l} \rho_{ijAl} \\ x_A^{(4)} &= \sum_{i,j,k} \rho_{ijkA}\end{aligned} \quad (20)$$

$x_A^{(s)}$ is also related to the site variables $\lambda^{(s)}$ by:

$$x_A^{(s)} = \frac{\lambda^{(s)}}{z_\alpha}\frac{\partial z_\alpha}{\partial \lambda^{(s)}} = \frac{\partial \ln z_\alpha}{\partial \ln \lambda^{(s)}} = \frac{k_B T \partial \ln z_\alpha}{k_B T \partial \ln \lambda^{(s)}} = -\frac{1}{m_\alpha N}\frac{\partial \Omega_\alpha}{\partial \mu_A^{(s)}} \quad (21)$$

The above relation reveals that the site fraction $x_n^{(s)}$ and site chemical potential $\mu_n^{(s)}$ are thermodynamic conjugate variables. The site variables $\lambda^{(s)}$ are Lagrange multipliers for the superposition relations in Eq.(20) [64].

Using these superposition relations from Eq.(20), the overall chemical potential of basic cluster $\alpha$ in Eq.(5) can be expressed in terms of $x_n^{(s)}$ and $\lambda_n^{(s)}$:

$$\begin{aligned}
\mu_\alpha &= \sum_c \rho_c \mu_\alpha(c) = \sum_{i,j,k,l} \rho_{ijkl}\mu_\alpha(ijkl)\\
&= \sum_{j,k,l}\rho_{Ajkl}\mu_A^{(1)} + \sum_{i,k,l}\rho_{iAkl}\mu_A^{(2)} + \sum_{i,j,l}\rho_{ijAl}\mu_A^{(3)} + \sum_{i,j,k}\rho_{ijkA}\mu_A^{(4)}\\
&= x_A^{(1)}\mu_A^{(1)} + x_A^{(2)}\mu_A^{(2)} + x_A^{(3)}\mu_A^{(3)} + x_A^{(4)}\mu_A^{(4)}\\
&= x_A^{(1)}k_B T \ln \lambda_A^{(1)} + x_A^{(2)}k_B T \ln \lambda_A^{(2)} + x_A^{(3)}k_B T \ln \lambda_A^{(3)} + x_A^{(4)}k_B T \ln \lambda_A^{(4)}\\
&= \sum_{s,n} x_n^{(s)} k_B T \ln \lambda_n^{(s)} \quad (22)
\end{aligned}$$

Finally, substituting Eqs.(4) and (22) into Eq.(3), the FYL-CVM free energy function as a function of $\lambda^{(s)}$ and $T$ is obtained in its most general form:

$$\begin{aligned}
\Delta F_{FYL-CVM}(\lambda^{(s)}, T) &= N\sum_{\alpha \in R} a_\alpha m_\alpha (\mu_\alpha + \Omega_\alpha)\\
&= Nk_B T \sum_{\alpha \in \{basic\ clusters\}} \left[a_\alpha m_\alpha \left(\sum_{s,n} x_n^{(s)}\ln\lambda_n^{(s)} - \ln z_\alpha\right)\right]\\
&\quad + Nk_B T \sum_{\alpha \in \{sub-clusters\}} \left[a_\alpha m_\alpha \left(\sum_c \rho_c \frac{\mu_\alpha(c)}{k_B T} - \ln z_\alpha\right)\right] \quad (23)
\end{aligned}$$

Note that the FYL transform is only carried for basic clusters, and the sub-clusters are expressed in these transformed clusters using the superposition relations. For a binary AB alloy on an fcc lattice with tetrahedron as the basic cluster, $\alpha$ can be tetrahedron, pair, or point clusters. The values of Möbius number $a_\alpha$, multiplicity $m_\alpha$, and possible sites $s$ depend on the cluster type and can be found from **Table 1** [57].

**Table 1.** The values of $a_\alpha$, $m_\alpha$, and $s$ for different cluster type $\alpha$ for an fcc AB alloy.

| Cluster type $\alpha$ | $a_\alpha$ | $m_\alpha$ | $s$ |
|---|---|---|---|
| Tetrahedron ($t$) | 1 | 2 | 1,2,3,4 |
| Pair ($p$) | -1 | 1 | 1,2 |
| Point ($a$) | 1 | $\frac{5}{4}$ | 1 |

Therefore, $\Delta F_{FYL-CVM}$ for an fcc AB alloy is expressed as

$$\Delta F_{FYL-CVM}(\lambda^{(s)}, T) = 2Nk_BT \left( \sum_{s=1}^{4} x_A^{(s)} \ln \lambda^{(s)} - \ln z_t \right)$$

$$-Nk_BT \left[ \sum_{\substack{s_1,s_2 \in \{i,j,k,l\} \\ s_1 \neq s_2}} \left( \sum_{i,j} \rho_{ij} \frac{\mu_\alpha(ij)}{k_BT} - \ln z_p \right) \right]$$

$$+ \frac{5Nk_BT}{4} \sum_s \left[ x_A^{(s)} \ln x_A^{(s)} + \left(1 - x_A^{(s)}\right) \ln\left(1 - x_A^{(s)}\right) - \ln z_a \right] \quad (24)$$

This equation applies to both ordered and disordered phases. Equilibrium is reached by varying the site variable $\lambda^{(s)}$ to minimize the total FYL-CVM free energy $\Delta F_{FYL-CVM}$. Like in CVM, the subclusters' probabilities are fully determined by the probabilities of basic clusters through the superposition relation [66]. Therefore, the pair clusters' probability $\rho_{ij}$ and site fractions $x_A^{(s)}$ in Eq.(24) are both dependent on $\lambda^{(s)}$. Only the effective energies for the basic cluster [50] will be used, leaving the pair and point cluster energies to be zero and $z_p = z_a = 1$. Therefore, the pair and point grand partition function terms $\ln z_p$ and $\ln z_a$ will vanish to zero.

2.4. *Comparison of CVM and FYL-CVM*

The key difference between conventional CVM and our FYL-CVM model is the use of different configurational variables: cluster variable $\rho_c$ in CVM vs. site variable $\lambda_n^{(s)}$ in FYL-CVM. We can relate these two models by replacing the CVM's cluster probabilities with the FYL-transformed cluster probabilities.

The CVM Helmholtz free energy in its general form can be expressed as [58,67,68]

$$\Delta F_{CVM} = \sum_\alpha \left[ a_\alpha m_\alpha N \left( \sum_c \rho_c \varepsilon_c + k_BT \sum_c \rho_c \ln \rho_c \right) \right] \quad (25)$$

where $c$ represents all the distinctive cluster configurations. Thus, comparing Eqs.(23) and (25), to prove $\Delta F_{FYL-CVM} = \Delta F_{CVM}$, we need to prove the following relation for the basic cluster:

$$k_B T \left( \sum_{s,n} x_n^{(s)} \ln \lambda_n^{(s)} - \ln z_\alpha \right) = \left( \sum_c \rho_c \varepsilon_c + k_B T \sum_c \rho_c \ln \rho_c \right) \tag{26}$$

Substituting the FYL-transformed cluster probability

$$\rho_c = \frac{\prod_{s,n} \lambda_n^{(s)} \exp\left(-\frac{\varepsilon_c}{k_B T}\right)}{z_\alpha} \tag{27}$$

into the RHS of Eq.(26), we have

$$\begin{aligned}
RHS &= \sum_c \rho_c (\varepsilon_c + k_B T \ln \rho_c) \\
&= \sum_c \rho_c \left[ \varepsilon_c + k_B T \ln \frac{\prod_{s,n} \lambda_n^{(s)} \exp\left(-\frac{\varepsilon_c}{k_B T}\right)}{z_\alpha} \right] \\
&= \sum_c \rho_c \left\{ \varepsilon_c + k_B T \left[ \left( \sum_s \ln \lambda_n^{(s)} \right) - \frac{\varepsilon_c}{k_B T} - \ln z_\alpha \right] \right\} \\
&= k_B T \left[ \sum_c \rho_c \left( \sum_s \ln \lambda_n^{(s)} \right) - \sum_c \rho_c \ln z_\alpha \right] \\
&= k_B T \left( \sum_{s,n} x_n^{(s)} \ln \lambda_n^{(s)} - \ln z_\alpha \right) = LHS
\end{aligned} \tag{28}$$

The final equality is based on the following two equalities. One is

$$\sum_c \rho_c \left( \sum_s \ln \lambda_n^{(s)} \right) = \sum_{s,n} x_n^{(s)} \ln \lambda_n^{(s)} \tag{29}$$

which is due to the superposition relation:

$$x_n^{(s)} = \sum_{c(s)} \rho_{c(s)} \tag{30}$$

where $c(s)$ represents the set of cluster configurations relevant to site $s$.

The other equality is the consequence of probability conservation

$$\sum_c \rho_c \ln z_\alpha = \ln z_\alpha \tag{31}$$

For the binary AB alloy on an fcc lattice, we have

$$\Delta F_{FYL-CVM}(\lambda^{(s)}, T) = 2Nk_BT\left(\sum_{s=1}^{4} x_A^{(s)} \ln \lambda^{(s)} - \ln z_\alpha\right)$$

$$-Nk_BT \sum_{i,j} \rho_{ij}\left(\varepsilon_{ij} + \ln \rho_{ij}\right)$$

$$+\frac{5Nk_BT}{4}\sum_s \left[x_A^{(s)} \ln x_A^{(s)} + \left(1 - x_A^{(s)}\right)\ln\left(1 - x_A^{(s)}\right)\right] \qquad (32)$$

$$\Delta F_{CVM}(\rho_{ijkl}, T) = 2Nk_BT \sum_{i,j,k,l} \rho_{ijkl}\left(\varepsilon_{ijkl} + \ln \rho_{ijkl}\right)$$

$$-Nk_BT \sum_{i,j} \rho_{ij}\left(\varepsilon_{ij} + \ln \rho_{ij}\right)$$

$$+\frac{5Nk_BT}{4}\sum_s \left[x_A^{(s)} \ln x_A^{(s)} + \left(1 - x_A^{(s)}\right)\ln\left(1 - x_A^{(s)}\right)\right] \qquad (33)$$

Compare Eqs. (32) and (33), we can see that the free energy functions of CVM and FYL-CVM are almost identical except the basic cluster terms with different configurational variables [64].

## 3. Application to the fcc AB prototype system

In this section, we apply our new FYL-CVM model to a hypothetical AB prototype system on an fcc lattice and compare our results with existing thermodynamic models (CVM, CSA, and MC). The tetrahedron is chosen as the basic cluster because it is the smallest three-dimensional structure to capture the symmetry of the ordering phases that we studied here: L1$_0$, L1$_2$, and disordered fcc A1 phase.

We set up the cluster energy of this prototype AB fcc system by taking the first nearest-neighbor pair energies as $J = \varepsilon_{AA} = \varepsilon_{BB} = k_B$ and $-J = \varepsilon_{AB} = \varepsilon_{BA} = -k_B$. Based on this pairwise energy setting, we calculate the tetrahedron cluster energy by counting the number of the bond pairs in the corresponding cluster, e.g., $\varepsilon_{AAAA} = \varepsilon_{BBBB} = 6J$, $\varepsilon_{AAAB} = \varepsilon_{AABA} = \varepsilon_{ABAA} = \varepsilon_{BAAA} = \varepsilon_{BBBA} = \varepsilon_{BBAB} = \varepsilon_{BABB} = \varepsilon_{ABBB} = 0$, $\varepsilon_{AABB} = \varepsilon_{ABAB} = \varepsilon_{BABA} = \varepsilon_{BBAA} = \varepsilon_{ABBA} = \varepsilon_{BAAB} = -2J$. However, the pairwise energy model suffers from the ignorance of multi-body interactions. The purpose of considering only pair interactions in this work is to showcase the baseline behavior of different models in a model fcc system. The energy model suitable for FYL-CVM will be discussed more in Section 4.1.

## 3.1. *Phase diagram*

The phase diagram is calculated in the chemical potential and temperature space. The phase boundaries between the different phases are determined when they share the same equilibrium grand potential [63]. The minimization of the FYL-CVM free energy is performed with the gradient descent (GD) method to calculate this potential. The GD method is a first-order iterative optimization algorithm for finding a local minimum of a differentiable function. Although the FYL-CVM free energy function is non-linear and non-convex [68] and the GD method only guarantees converging to the local minima, it still converges to the proper global minima in most cases based on our tests. The GD method and its variant, stochastic GD method, recently is proved to be helpful to avoid saddle points efficiently [69]. The property of the cluster grand partition function (Eq.(14)) from FYL-CVM guarantees the normalization constraint of cluster variables to be automatically satisfied. Therefore, the site variables are like correlation functions, which constitute an independent set of variables. Unlike in the Natural Iteration Method [70], the Lagrange multiplier method is not used in the minimization of FYL-CVM free energy here. The use of FYL-CVM may also enable the application of Newton-Raphson method for minimization without correlation functions [62]. A detailed discussion of the free energy minimization algorithms will be presented in our future work.

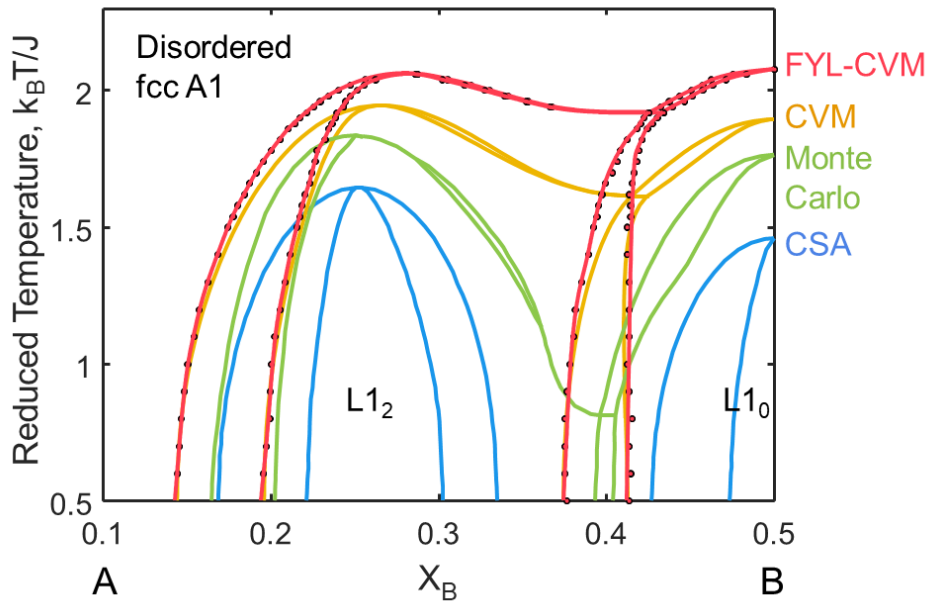

**Figure 3.** Calculated prototype phase diagrams of AB binary alloy on fcc lattice by 4 different models: FYL-CVM in current work (represented with dot-line), CVM [70,71], Monte Carlo [72,73], and CSA [28].

The calculated FYL-CVM phase diagram along with CVM, CSA, and MC ones are presented in **Figure 3**. The MC calculated results is based on the Ising model from Inden's work [72,73] which serves as a benchmark for comparison. The phase diagram from CVM is adopted from [70,71]. The phase diagram of CSA is adapted from Li's original work [28] because CSA is fundamentally the same as Yang and Li's generalized quasi-chemical approximation [37,38]. All the energy settings in different models including the MC method are consistent with each other. The dots represent the equilibrium compositions of the phases and are connected by fitting appropriate polynomial. The reduced order-disorder transition temperatures ($k_B T/J$) between the $L1_0/L1_2$ phase and disordered phase from different models are shown in Table 2.

Taking the MC results as the benchmark, our FYL-CVM model correctly describes the key topology of the phase diagram with all the necessary phases. The calculated phase diagram from FYL-CVM nearly replicates that from CVM, except for slight overestimation in the transition temperatures. This indicates that FYL-CVM, while utilizing significantly less variables, preserves the essential feature of CVM. The diagram from CSA misses the critical invariant temperature, and the disordered phase becomes stable to low temperatures. This indicates a large error and CSA fails to reproduce the correct topology of the fcc order-disorder phase diagram.

**Table 2.** Order-disorder transition temperatures in different thermodynamic models for the prototype fcc AB system.

| Thermodynamic Model | $L1_2$ | $L1_0$ |
|---|---|---|
| Bragg-Williams [74] | 3.28 | 4 |
| Bethe-Peierls [75] | 1.78 | - |
| Quasi-Chemical [76] | - | 3.57 |
| CSA [38] | 1.64 | 1.46 |
| FYL-CVM (this work) | 2.06 | 2.08 |
| CVM [70] | 1.94 | 1.89 |
| Monte Carlo [77] | 1.87 | 1.76 |
| Monte Carlo [78] | 1.84 | 1.76 |

The results from the mean-field Bragg-Williams model (i.e., the sublattice model or the compound energy formalism [23]) used in conventional CALPHAD are presented in **Figure 4**. As shown in **Figure 4**, the phase diagram from sublattice model is inferior to those from the cluster-based models, which better accounts for the configurational entropy, in terms of the predicted diagram topology and transition temperature predictions. The sublattice model significantly overestimates the single congruent transition temperature at $X_B=0.5$ which all the ordered phases falsely share. This is because CSRO is missing in the sublattice model in conventional CALPHAD, but it significantly influences the phase diagram topology and the transition temperatures. The above comparison indicates the necessity of developing a cluster-based closed-form analytical model for CALPHAD.

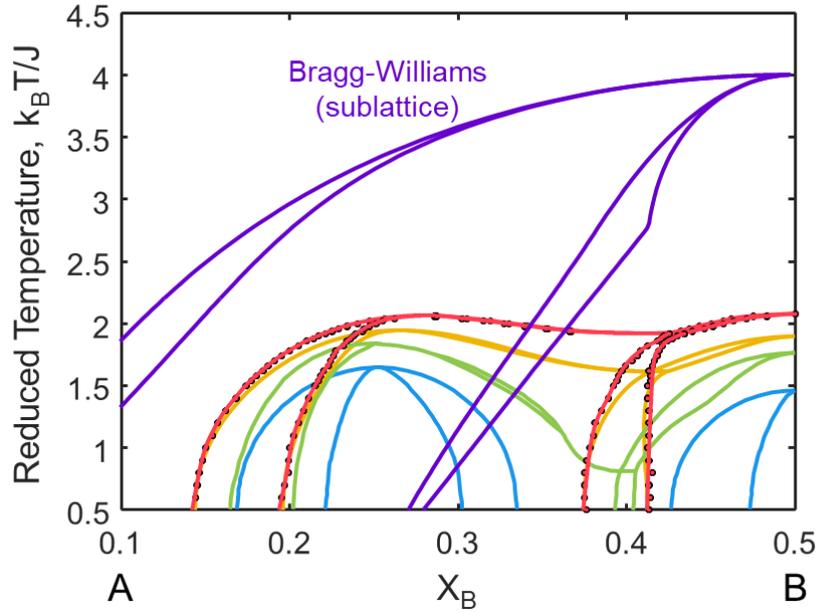

**Figure 4.** Prototype ordering phase diagrams of AB binary alloy on fcc lattice in comparison with the Bragg-Williams (sublattice) model in purple color, which is adapted from [79] and the data is reproduced with Thermo-Calc2023b [80]. The phase diagrams plotted in other colors are from the cluster-based models indicated in Figure 3.

The predicted invariant transition temperature (triple point) from FYL-CVM has sizable difference compared with that from MC in **Figure 3**. It needs to be mentioned that locating the invariant temperature from MC is non-trivial due to the frustration effect of the fcc lattice [81]. The fluctuating antiphase boundaries (APB) [82,83] appeared in MC simulations also make it harder to interpret the results. Consequently, there were controversies [77,78,83] in the literature regarding where this invariant point should be. Nonetheless, the MC predicted values of invariant transition temperature in the literature converged to $k_BT/J \approx 1$ while CVM with tetrahedron approximation to $k_BT/J \approx 1.6$ [83] and FYL-CVM to $k_BT/J \approx 2$ in the current work. Using the Tetrahedron-Octahedron approximation of the CVM (TO-CVM), which includes the 2nd nearest-neighbor interaction, can further bring down the invariant temperature to $k_BT/J \approx 1.5$ [84]. The computational difficulty of using a bigger basic cluster in TO-CVM can be significantly alleviated by FYL-CVM. In the modeling of real alloy systems, this intrinsic error from FYL-CVM can be further diluted through proper parameterization.

### 3.2. *Thermodynamic properties*

The difference in the calculated phase diagrams can be rationalized by the thermodynamic properties calculated from different models. All calculated results (entropy S, energy E, free energy F, and heat capacity $C_V$) for an $A_{0.5}B_{0.5}$ alloy on a fcc lattice are shown in **Figure 5**. The CSA and FYL-CVM thermodynamic data are calculated by the authors, with others reproduced from literature. The CSA data calculated here agree with the previous results by Oates [37,38]. The heat capacity $C_V$ is calculated by taking the derivative of the internal energy with respect to temperature.

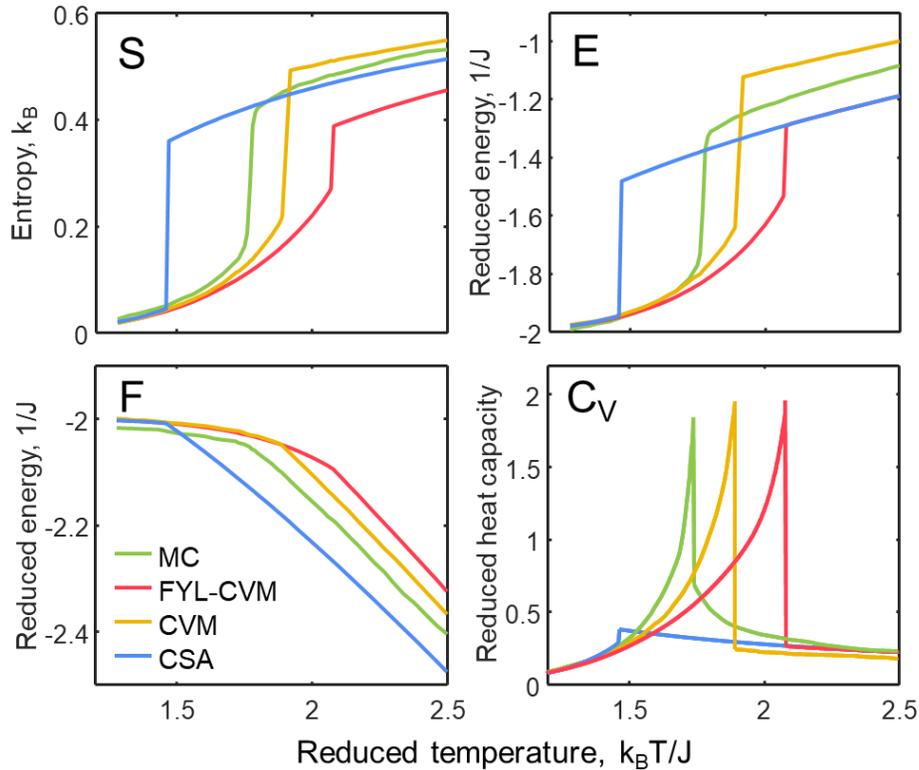

**Figure 5.** Calculated S, E, F, and $C_V$ for an $A_{0.5}B_{0.5}$ alloy for $L1_0$ and A1 phases by 4 different models. The CVM data is based on [70,71,85]; The CSA data is recalculated following [37,38]; The MC data is considered from [86,87].

Benchmarked against the MC results, CVM delivers the highest accuracy among all analytical models. The thermodynamic properties for ordered $L1_0$ phase from all models are in good agreement while differences appear for the disordered fcc A1 phase. CVM overcalculates the entropy and energy while FYL-CVM undercalculates them at high temperatures consistently. The underestimation of entropy from FYL-CVM is due to the FYL transform which compresses the configurational space and thus lowers the entropy, as evident from Figure 5. CSA shows the poorest performance among all models with a significant underestimation of the order-disorder transition temperature as well as a very small "$\lambda$ peak" in heat capacity. The energy curves of FYL-CVM and CSA almost overlap at high temperatures. This is because the basic cluster probability distributions of both FYL-CVM and CSA at high temperatures are the same after the FYL transform, different from that of CVM without the FYL transform. Another interesting observation is that CSA as well as CVM produces the best entropy prediction comparable with the MC entropy for the disordered phase.

It is reported that CVM produces correlation functions that are too close to zero, which leads to an overestimation of the energy, E, and at the same time, to an underestimation of -TS, so the free energy F=E-TS is more accurate than either of its parts [88]. This is known as the error cancellation effect of CVM. The calculated errors of thermodynamic properties (F, E, and -TS) between the MC results and all analytical models are shown in **Figure 6**. We can see that the error cancellation effect exists in all cluster-based models. It is most prominent above the order-disorder transition temperature for CVM and FYL-CVM while below the transition temperature for CSA. The error cancellation behavior of FYL-CVM is similar to that of CVM but larger in extent. This effect implies that it is probably better to holistically improve the CVM model rather than optimize its energy and entropy terms separately.

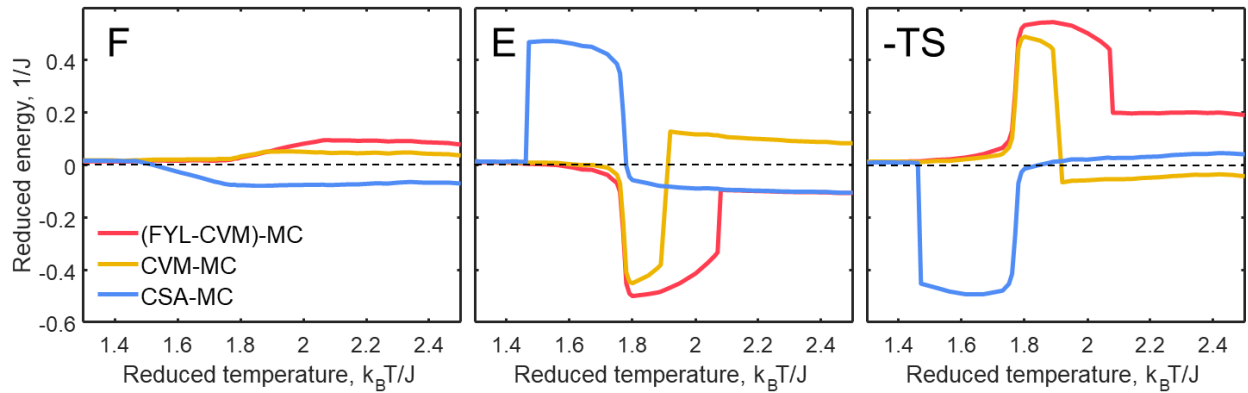

**Figure 6.** F, E, and -TS errors as a function of reduced temperature for an $A_{0.5}B_{0.5}$ alloy from the FYL-CVM, CVM, and CSA models in comparison with the MC results showing the error cancellation effect.

The thermodynamic signatures of order-disorder transition deserve to be discussed more. Generally, the order-disorder transition in alloys is a second-order transition [89,90]. The latent heat (first derivative of free energy) as a function of temperature should not have any discontinuity but the heat capacity (second derivative of free energy). However, here every cluster-based analytical model incorrectly produces the discontinuities in the latent heat plot. The energy discontinuity originates from the limit in correlation length in these cluster-based models, in which the basic clusters must be truncated at a certain range outside of which mean-field approximation is applied. This limited correlation length makes CVM, CSA and FYL-CVM perform poorly near the critical point where the correlation length is infinite [89]. Because the overall behavior of the phase diagram and thermodynamic properties is the most important, this deficiency of cluster-based models near the critical point is acceptable.

3.3. *Order parameters*

The calculated Warren-Cowley Short-Range Order (SRO) parameters $\alpha_{AB}$ and the Long-Range Order (LRO) parameters $\eta$ for the $A_{0.5}B_{0.5}$ fcc alloy from different models are shown in **Figure 7**.

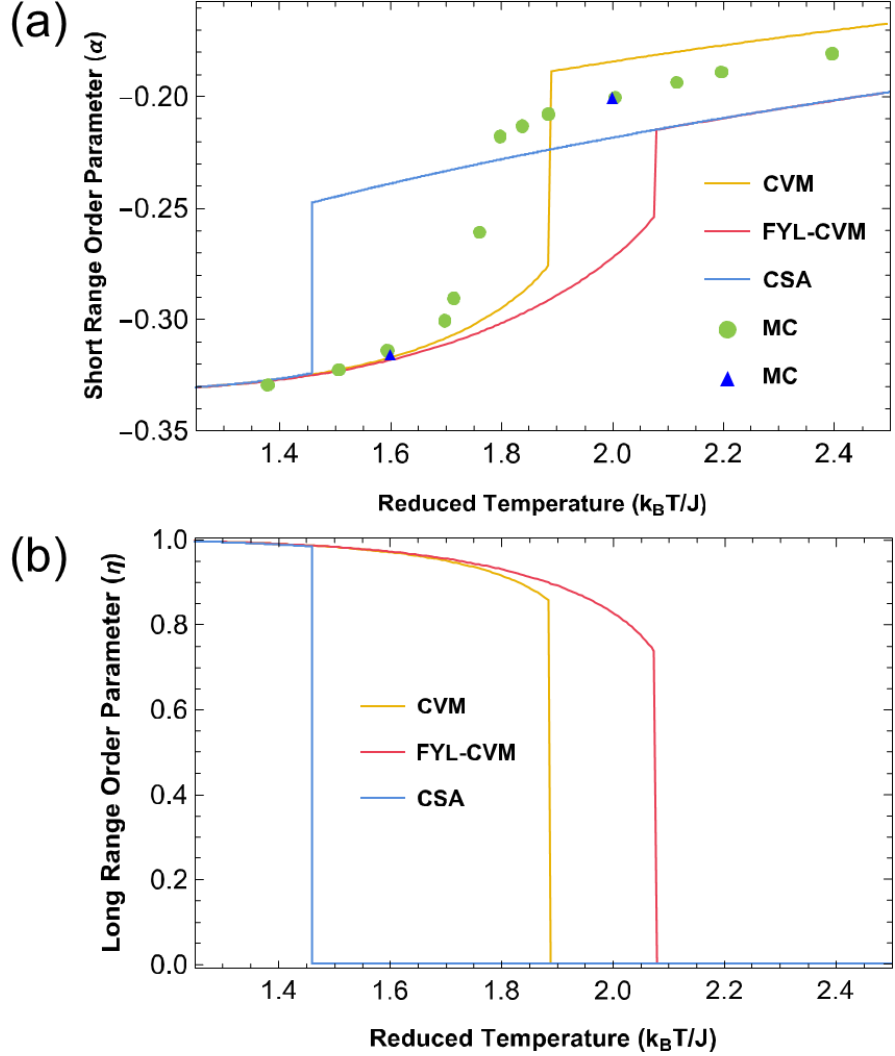

**Figure 7.** For the $A_{0.5}B_{0.5}$ fcc alloy: (a) Calculated Warren-Cowley SRO parameter $\alpha$ from CVM, FYL-CVM, CSA, and MC (●[91],▲[77]) models. (b) Calculated LRO parameters $\eta$ from CVM, FYL-CVM, and CSA.

The Warren-Cowley SRO parameter $\alpha_{AB}$ is calculated using [92,93]

$$\alpha_{AB} = 1 - \frac{\bar{\rho}_{AB}}{2x_A x_B} \qquad (34)$$

in which $\bar{\rho}_{AB}$ represents the overall cluster probabilities of nearest-neighbor AB pair for the tetrahedron cluster, $\bar{\rho}_{AB} = \rho_{AB} + \rho_{BA}$. $x_A$ and $x_B$ are compositions of the system.

The comparison of $\alpha_{AB}$ calculated for FYL-CVM using Eq. (34) along with the other thermodynamic models (CVM, CSA, and MC simulations [77,91]) is shown in **Figure 7**. The $\alpha_{AB}$ calculated using FYL-CVM for both L1$_0$ and A1 phases shows the same trend with the CSA,

CVM, and MC results. For the ordered phase at low temperatures, the $\alpha_{AB}$ calculated by all the cluster models agree well with each other and the departure from MC results is negligible. However, the high temperature behavior of $\alpha_{AB}$ in disordered phases is different between models with different order-disorder transition temperatures. For the disordered phase at high temperature, FYL-CVM, along with CSA underestimates the value of $\alpha_{AB}$ compared with MC while CVM overestimates it. This is very similar to the behavior of calculated entropies.

For the ordered L1$_0$ phase shown in **Figure 9**, the sublattices that are preferentially occupied by A atoms are represented as $\alpha$ sublattice while those by B atoms are $\beta$ sublattice. The average of the composition on each sublattice is made related to the average composition of the system. For the L1$_0$ phase, it can be expressed as

$$x_B = \frac{2x_B^\alpha + 2x_B^\beta}{4} \tag{35}$$

The LRO parameter $\eta$ is defined as the difference between the occupancy of the right component on the chosen sublattice to the occupancy of the same component on the other sublattice. For the L1$_0$ phase,

$$\eta = x_B^\beta - x_B^\alpha = x_A^\alpha - x_A^\beta \tag{36}$$

It can be observed from **Figure 7** that the calculated LRO parameter from FYL-CVM approaches the limiting value of 1 as the temperature approaches 0K, similar to that of CSA and CVM. The incorrect discontinuity shown in **Figure 7** is due to the limited correlation lengths in cluster models as discussed in Section 3.2. Overall, the results show that both the SRO and LRO parameters could be calculated reasonably well with FYL-CVM.

## 4. Discussion

### 4.1. *The energy model*

The FYL-CVM model still requires the basic cluster energies in Eq. (15) as the input model parameters(15). The FYL transform reduces the number of minimizing variables, but not model parameters. Kikuchi only applied the CVM hierarchical expansion to the configurational entropy but not the energy [50,66]. In FYL-CVM, we follow Kikuchi's approach by expressing total energy using the effective basic cluster energy terms without any sub-clusters:

$$E = \sum_{i,j,k,l} \rho_{ijkl}\varepsilon_{ijkl} \tag{37}$$

where $\rho_{ijkl}$ is the probability of the cluster $ijkl$. $\varepsilon_{ijkl}$ is the corresponding effective cluster energy and can be obtained from first-principles calculations or experimental data or a combination of both. We will discuss the approach to determine the cluster energies based on first-principles calculations in our forthcoming publications.

There was a prevalent use of pairwise bond energy terms in the early CVM literature [50,66,94,95] due to limited computational power. However, it has limited accuracy due to ignoring the multi-body interactions like those encapsulated in a tetrahedron cluster. The pairwise bond energy model for fcc AB alloy is used in current work only to demonstrate the baseline behavior of different models. Adding multi-body interactions will likely improve the comparison of the phase diagram from CVM or FYL-CVM with that from MC, which is suggested to be used in modeling real systems.

Recent CVM calculations [94,96] model the energy using another popular alloy theoretical technique Cluster Expansion (CE) [62,97]. CE expands the energy in terms of the effective cluster interaction (ECI) and orthonormal correlation functions in a generalized Ising model [98]:

$$E = \sum_\alpha m_\alpha J_\alpha \xi_\alpha(\vec{\sigma}) = J_0 + \sum_i J_i \sigma_i + \frac{1}{2}\sum_{i \neq j} J_{ij}\sigma_i\sigma_j + \frac{1}{3!}\sum_{i \neq j \neq k} J_{ijk}\sigma_i\sigma_j\sigma_k + \cdots \tag{38}$$

where $m_\alpha$ is the multiplicity of cluster $\alpha$, $J_\alpha$ is the ECI, and $\xi_\alpha(\vec{\sigma})$ is the correlation function. CE captures the interactions beyond the pairwise nearest neighbor. The use of CE for the energy model is well established and often coupled with MC to calculate phase diagrams [62,99–101]. The CE energy model was demonstrated in a previous CVM-CALPHAD type framework [102,103]. However, the ECIs are not combinable due to the orthonormality of correlation functions, which makes it impossible to reuse the descriptions of the constituent lower order sub-systems in a higher order description. This disqualifies CE for large databases development for multicomponent systems. There were attempts to simplify CE to make it compatible with CALPHAD by developing approximate analytical solutions for correlation functions [104–106]. However, the scalability of this approach is not yet reported. The combinability of the effective basic cluster energy model (Eq.(37)) is the main reason for us to use it in FYL-CVM. In Supplementary Materials, we attempt to derive an explicit relationship between site variables $\lambda$ in FYL-CVM and correlation functions $\xi$ in CE but there is none.

4.2. *Cluster interference vs. non-interference*

The FYL transform is a fundamental ansatz in the original quasi-chemical model by Fowler and Guggenheim [107] and the generalized quasi-chemical model by Yang and Li [108]. The name "quasi-chemical" comes from the similarity to the mass action relation in chemical reaction equilibrium. The original quasi-chemical model is limited to pairs which are naturally of non-interference. Yang and Li generalized it to clusters and kept the non-interference assumption, which means clusters are independent. Our FYL-CVM model keeps the essence of the quasi-chemical model by adopting the mass-action-like FYL transform but combines it with Kikuchi's CVM which has a rigorous treatment of the interference effect. The accuracy lost due to FYL transform is simply due to the compression of the internal cluster variable space, not from the atomic correlations described by the CVM hierarchical formula.

The CSA [37,38] revived by Oates is attributed to Yang's original work [25,26] which has the FYL transform [46]. The CSA entropy is a two-term entropy which only contains contributions from the basic clusters and points [37]. It is obtained if one considers the mixing of clusters which are not permitted to share edges or bonds, also known as the non-interference assumption. This leads to significant inaccuracy as demonstrated in Section 3. The main difference between FYL-CVM and CSA [37] is whether the non-interference assumption [25] is adopted, as illustrated in **Figure 8**. The non-interference assumption implies that sub-clusters are not allowed and ignores the contribution from overlapping clusters. It was used to simplify the treatment of cluster correlations before Kikuchi's CVM [29] was invented. By taking the CVM hierarchical formalism, FYL-CVM model allows the overlapping between the clusters and quantifies the overlapping using inclusion-exclusion principle [59] so that the accuracy is much improved compared with CSA. Oates later modified CSA [38] by adopting one additional modeling parameter to mimic the overlapping interactions albeit it still uses the incorrect non-interference assumption. The *ad hoc* modeling parameter in modified CSA is unphysical, and it is difficult to be extended into multicomponent systems.

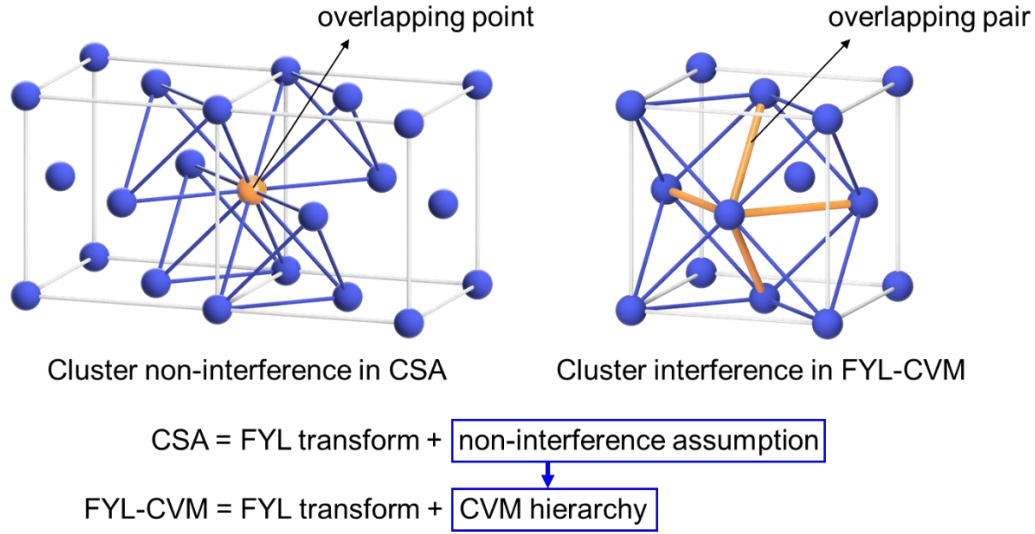

**Figure 8.** Illustration of the non-interference vs. interference of tetrahedron clusters on the fcc lattice considered in CSA and CVM, respectively. The overlapping point in CSA and overlapping pair in FYL-CVM are highlighted in orange.

4.3. *Advantages of FYL-CVM*

Through the FYL-CVM model, we successfully incorporate intrinsic CSRO into CALPHAD, while maintaining its practicality and efficiency. It leverages statistical mechanics to yield a more physical description of configurational entropy and opens the door to cluster-based CALPHAD database development. The advantages to be gained in introducing more physics into solution models are obvious: correct phase diagram topology, correct lattice structures, correct order-disorder behavior, etc. The recognition of these properties in the models should give more confidence in data extrapolation and involve fewer fitting parameters, which is the subject of our subsequent publications. Additionally, FYL-CVM also offers superior computational efficiency (to be quantified) and scalability compared to CVM. The minimizing variables are reduced from the order of $n^s$ in CVM to the order of $n \times s$ in FYL-CVM, which makes it feasible to extend FYL-CVM to multicomponent systems. Fundamentally the key difference between the models is about free energy minimization in different spaces: Bragg-Williams (sublattice) model in the (T, *x*) space, CVM in the (T, $\rho$) space, and FYL-CVM in the (T, $\lambda$) space.

The use of FYL-CVM could potentially bring many benefits to the CALPHAD modeling, such as avoiding spurious ordering caused by the unphysical ordering behavior of sublattice model, a more convenient model for vacancy thermodynamics, and removal of hypothetical endmembers [109]. The configurational and non-configurational (vibrational, elastic, electronic) contributions to free energy will be modeled separately, gaining insights into their respective effects on phase stability. The non-configurational free energy contributions will be discussed in a separate future paper. The versatility of the new solution model means it is applicable to metallic, ionic, and semiconductor alloys while offering scalability towards larger clusters and/or more components within the CVM hierarchy. This cluster-based free energy model is naturally more suitable for describing thermodynamics of coherent interfaces and inhomogeneous systems, like the anti-phase boundary energy [51] or the Cahn-Hilliard gradient energy term [110]. Understanding CSRO has important implications for the study of nucleation in solid solutions [111,112] because of its impact on coherent interfacial energy [113].

The FYL-CVM model includes the critical crystal structure information which is nearly absent in sublattice models. FYL-CVM also provides a unified description of order-disorder free energy instead of separating it into ordered and disordered parts as in the conventional CALPHAD. In CVM, sublattices are still needed to describe the LRO [114]. While in FYL-CVM, LRO can be inferred by observing the symmetry of site variables $\lambda^{(s)}$ after minimization. The symmetry of $\lambda^{(s)}$ means whether they are equal or not. For example, as shown in **Figure 9**, the tetrahedron in disordered fcc A1 phase would have $\lambda^{(1)} = \lambda^{(2)} = \lambda^{(3)} = \lambda^{(4)}$ (or equivalently $\mu^{(1)} = \mu^{(2)} = \mu^{(3)} = \mu^{(4)}$), but $\lambda^{(1)} \neq \lambda^{(2)} = \lambda^{(3)} = \lambda^{(4)}$ and $\lambda^{(1)} \neq \lambda^{(2)} = \lambda^{(3)} = \lambda^{(4)}$ in ordered L1$_2$ and L1$_0$ phases, respectively. The site variables naturally serve as the symmetry-adapted order parameters [115], which can treat ordered and disordered phases in a unified framework. A direct minimization of the FYL-CVM free energy function with site variables should reveal the order-disorder phase transitions. However, one must be careful in selecting an appropriate basic cluster so that it can represent all possible ordering structures of the crystal, that is equivalent to requesting the basic cluster to be large enough. A remark should be made on the choice of the optimum basic cluster. Choice of the basic cluster should ideally include the longest interaction distance [62]. But in practice it is sufficient to just make the basic cluster include all the major ordering types shown in the phase diagram, such as L1$_2$ and L1$_0$ for the fcc phase. The basic cluster size needs not to be equal to the domain size observed in experiments but should capture the unit cell size of ordering type exhibited in the ordered domain.

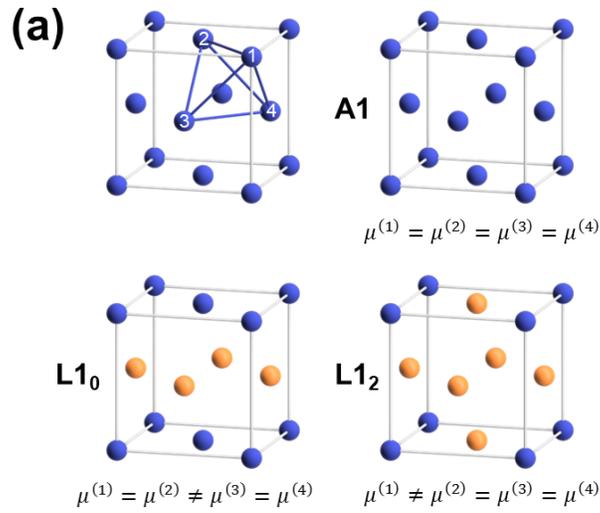

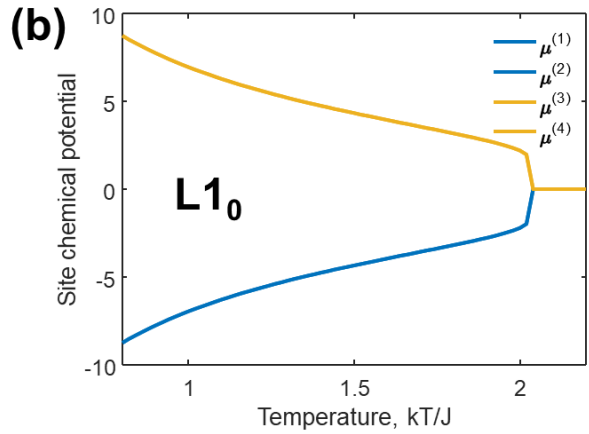

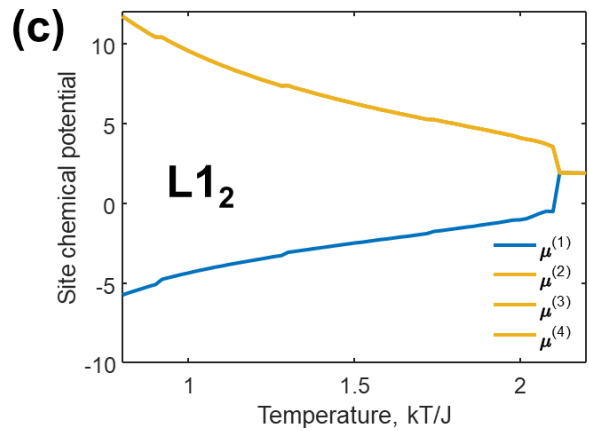

**Figure 9.** Site variables (site chemical potentials $\mu^{(s)}$) as a function of temperature as the order parameters of phases at (b) $A_{0.75}B_{0.25}$ ($L1_0$) and (c) $A_{0.5}B_{0.5}$ ($L1_2$) compositions.

One issue of the conventional CALPHAD method is that, being based on classical thermodynamics, it misses all the metastable, microstructural constituents, and transient phases that are at the core of the most successful engineering alloys [2]. CVM-CALPHAD will unleash the potential of CALPHAD to model these metastable and transient structures/phases. Our novel FYL-CVM model developed here is able to describe simultaneous ordering and clustering (phase separation) reactions by capturing the nuances of the curvature in free energy vs. composition (G-*x*) diagrams, such as those in the technologically important Ni-Al, Ni-Ti, and Al-Li alloys identified by Soffa and Laughlin [116].

Although CVM-CALPHAD frameworks with similar scope were proposed in previous literature [94,102], it was still considered too complicated for the multicomponent alloys at that time. After the detailed benchmark tests on the prototype AB system, we compare the performance traits to satisfy the CALPHAD requirements in **Table 3** to demonstrate FYL-CVM's key features compared to other solution models. We can see that FYL-CVM is the only model which possesses all the favorable traits to be compatible with CALPHAD, while achieving a good balance between accuracy and computational cost.

**Table 3.** Comparison of the expected performance of different analytical solution models on a low-medium-high scale, with more atomic correlations included from left to right.

|  | **Sublattice** | **CSA** | **FYL-CVM** | **CVM** |
|---|---|---|---|---|
| Include intrinsic CSRO? | No | Yes | Yes | Yes |
| Scalable to multicomponent? | Yes | Yes | Yes | No |
| Number of variables | Medium | Low | Low | High |
| Accuracy | Low for CSRO | Low | High | High |
| Computational cost | Low | Medium | Medium | High |
| Usability/Versatility | High | Low | Medium | Low |

## 5. Conclusion and outlook

In this work, we have developed a novel solution model, FYL-CVM, for the configurational contribution of the solid solution with intrinsic CSRO by applying the FYL transform to CVM. It is fundamentally a reduced order model for CVM. The number of variables with respect to which the free energy to be minimized is significantly lower, from the order of $n^s$ to $n \times s$, which makes it easy to extend to multicomponent alloys. Consequently, FYL-CVM satisfies all the requirements to be compatible with the CALPHAD framework and maintains a balance between accuracy and computational cost. It can calculate the phase diagram with correct topology and reasonably good thermodynamic properties for the fcc AB prototype system.

The hybrid CVM-CALPHAD framework enabled by FYL-CVM represents a new methodology for thermodynamic modeling that enables atomic-scale order to be exploited for materials design [117], which potentially leads to novel CCAs or other alloys. The FYL-CVM model can inspire new free energy models with SRO of magnetic spins [118], and similar methods could be extended to the realm of kinetics in analogy to the path probability method [119], both in the context of CALPHAD. It will open the door to cluster-based CALPHAD database development and even to thermodynamic databases of interfaces that are inputs for the phase-field simulations. Understanding CSRO also has important implications for the study of nucleation in solid solutions [111,112] where it may serve as the precursors for nucleation.

Numerous questions remain to be addressed to fully realize the vision of a CVM-CALPHAD thermodynamic framework. Such as: How to model the non-configurational contributions (elastic, vibrational, electronic) to free energy in a cost-effective way? How to parameterize the effective basic cluster energies in FYL-CVM accurately and efficiently for real alloy systems using experimental and theoretical data (i.e., cluster-based CALPHAD database development)? How to choose or design more efficient algorithms for minimizing the non-linear and non-convex FYL-CVM free energy function? The limitations of the FYL-CVM model need to be further explored, especially in modeling multicomponent systems. We aim to address these questions in our forthcoming publications, along with ongoing development of open-sourced codes for the CVM-CALPHAD method.

## Data Availability Statement

The datasets generated during the current study are available from the corresponding author (bicheng.zhou@virginia.edu) upon reasonable request.

## Declaration of Competing Interest

The authors declare that they have no known competing financial interests or personal relationships that could have appeared to influence the work reported in this paper.

## Acknowledgements

The present work was funded by the US National Science Foundation (NSF) through the CAREER Grant No. 2042284 as well as the support from the University of Virginia start-up funds.

## Appendix

List of acronyms used throughout this paper:

| Acronym | Definition |
|---|---|
| CSRO | Chemical Short-Range Order |
| LRO | Long-Range Order |
| CCA | Complex Concentrated Alloys |
| CALPHAD | CALculation of PHAse Diagrams |
| CVM | Cluster Variation Method |
| FYL | Fowler-Yang-Li |
| CSA | Cluster/Site Approximation |
| MC | Monte Carlo |
| CE | Cluster Expansion |
| ECI | Effective Cluster Interaction |
| GD | Gradient Descent |

List of symbols and their definitions in this paper:

| Symbol | Definition |
|---|---|
| $i, j, k, l$ | the species occupying lattice points in a cluster |
| $ijkl$ | the collection set of basic cluster configurations |
| $\varepsilon_{ijkl}$ | energy of clusters with configuration $ijkl$ |
| $c$ | cluster configuration |
| $\rho_{ijkl}, \rho_c$ | cluster probability of the clusters having configurations, $ijkl$ or $c$ |
| $\alpha_{AB}$ | Warren-Cowley short-range order parameter for a pair AB |
| $\eta$ | long-range order parameter |
| $m_\alpha$ | multiplicity of cluster $\alpha$ |
| $a_\alpha$ | Möbius number of cluster $\alpha$ |
| $\gamma_\alpha$ | Kikuchi-Barker coefficient of cluster $\alpha$ |
| $\sigma_i$ | spin variable of species $i$ |
| $J_\alpha$ | effective cluster interaction (ECI) |
| $\xi_\alpha(\vec{\sigma})$ | correlation function |
| $z_\alpha$ | grand partition function of cluster $\alpha$ |
| $\Omega$ | grand potential |
| $\Omega_\alpha$ | grand potential of cluster $\alpha$ |
| $\mu$ | chemical potential |
| $\mu_\alpha$ | chemical potential of cluster $\alpha$ |
| $\lambda_n^{(s)}$ | site variable/activity for component $n$ on site $s$ |
| $x_n^{(s)}$ | site mole fraction for component $n$ on site $s$ |

# SUPPLEMENTARY MATERIALS

# A Cluster-Based Computational Thermodynamics Framework with Intrinsic Chemical Short-Range Order: Part I. Configurational Contribution

Chu-Liang Fu, Rajendra Prasad Gorrey, Bi-Cheng Zhou[*]

Department of Materials Science and Engineering, University of Virginia, Charlottesville, Virginia 22904, USA

## S1. FYL-CVM derivation using canonical ensemble

### S1.1 Yang's original derivation

In this section, we briefly go through Yang's original derivation [1,2] of the Yang-Li approximation (or equivalently CSA). We follow Yang's original approach to derive the FYL-CVM.

Based on statistical mechanics of the canonical ensemble, we have

$$F(\vec{x}, T) = -k_B T \ln Z(\vec{x}, T) \tag{39}$$

$\vec{x}$ is the fraction vector for the multicomponent alloys, $\vec{x} = (x_i) = (x_1, x_2, x_3, \ldots, 1 - \sum_i x_i)$. We define the number of the components is $n$. So $\vec{x} = (x_1, x_2, x_3, \ldots, x_n)$. However, the previous expression provides the constraint directly and saves the number of variables. $T$ represents the temperature. $Z$ is the canonical partition function, which is different from the grand canonical partition function $\mathcal{Z}$. $k_B$ represents the Boltzmann constant.

We have the relation between canonical partition function and internal energy:

$$\ln Z = \ln Z(x, \infty) + \int_\infty^T \frac{E}{k_B T^2} dT \tag{40}$$

where $Z(x, \infty)$ is the high temperature limit for the canonical partition function

$$Z(x, \infty) = \prod_s \left( \frac{\left(\frac{N}{S}\right)!}{\prod_n \left(x_n^{(s)} N\right)! \left[\left(1 - \sum_{s,n} x_n^{(s)}\right) N\right]!} \right) \tag{41}$$

[*] Corresponding author: Bi-Cheng Zhou (bicheng.zhou@virginia.edu)

$x_n^{(s)}$ is the site probability of the component $n$ for site $s$ relative to the selected basis cluster and $x_s = \frac{1}{S}\left(\sum_n x_n^{(s)}\right)$ is the composition of chemical element $n$. $S$ is the number of sites in a basis cluster. $N$ is the total number of atoms. The use of the prefactor is to keep mass conservation.

In the thermodynamic limit, applying the Stirling's formula we have:

$$\ln Z(x, \infty) = -\frac{N}{S} \sum_{s,n} x_n^{(s)} \ln x_n^{(s)} \tag{42}$$

The main difficulty is how to treat the integral in Eq. (40). Yang used the Legendre transform trick [1,2]. First, we express the energy term with the CVM-like cluster formalism to capture the short-range order [3–5]:

$$E = N \sum_{\alpha \in R} a_\alpha m_\alpha \varepsilon_\alpha = N \sum_{\alpha \in R} a_\alpha m_\alpha \sum_{i,j,k,l} \rho_{ijkl} \varepsilon_{ijkl} \tag{43}$$

This CVM-like energy expansion was originally developed by Kikuchi to quantify the entropy hierarchically with the cluster & sub-cluster relation. $a_\alpha$ and $m_\alpha$ are the coefficients to eliminate the overcounting in various clusters' contributions [6,7]. $\rho_{ijkl}$ represents the probability of the specific type of the clusters $\alpha$, $\varepsilon_{ijkl}$ represents the corresponding cluster energy with a specific decoration, and $\varepsilon_\alpha$ is the averaged energy for cluster $\alpha$. $ijkl$ is the label notation to label the different types of the possible decoration of such cluster $\alpha$. $\alpha$ is the notation to represent different types of clusters.

This expression provides the interacting part of the internal energy. The CSRO is inserted by the different types of clusters. The cluster probability provides the change to use statistical description of the nature of the ordering. The randomly mixed clusters with CSRO constitute the whole ensemble in the crystal.

We represent the probability as:

$$\rho_{ijkl} = \frac{\left(\prod_{s=1}^{S} \xi_s(ijkl)\right) \exp\left(-\frac{\varepsilon_{ijkl}}{k_B T}\right)}{z_\alpha}; \quad z_\alpha = \sum_{i,j,k,l} \left(\prod_{s=1}^{S} \xi_s(ijkl)\right) \exp\left(-\frac{\varepsilon_{ijkl}}{k_B T}\right) \tag{44}$$

with $\xi_s(t) = \lambda_n^{(s)}$. Here $\xi_s$ is a function to represent the site contribution to the probability to be some component, the contribution is quantified by the parameter $\lambda_n^{(s)}$ which is determined by the site relative to the cluster and the component. The reason why the final condition exists is we can use this setting to set up the reference value of these parameters to save the number of the parameters. It is similar to set up the zero point of the potential energy. $\varepsilon_{ijkl}$ is the corresponding cluster energy. $z_\alpha$ is the grand partition function of cluster $\alpha$.

Here we have already performed the mean-field approximation among the cluster and the FYL-transform. The difference between grand canonical ensemble method developed in the current work and the canonical ensemble method originated from Yang and generalized by us here is we can only understand such variable $\lambda$ as the statistical weighted factor between the different microstates to adjust the distribution of different cluster to approach the minimization of the free energy to calculate the equilibrium state. However, grand canonical ensemble can provide much more clear physical meaning of these variables.

To avoid the integral in the partition function and average energy relation, we follow the Yang's idea to perform the Legendre transform of the partition function:

$$\ln z_\alpha \left(\ln \lambda_n^{(s)}, \ldots, T\right) \to \Psi_\alpha\left(x_n^{(s)}, \ldots, T\right) \tag{45}$$

where $z_\alpha$ is the grand canonical partition function of cluster. Note that there is no canonical cluster partition function $z_\alpha$ as each cluster is an open system where the cluster chemical potential is subject to variation. This implies

$$\Psi_\alpha\left(x_n^{(s)}, \ldots, T\right) = \ln z_\alpha - \sum_{s,n} x_n^{(s)} \ln \lambda_n^{(s)} \tag{46}$$

The key relation to prepare for this Legendre transform is based on

$$\frac{\partial \ln z_\alpha}{\partial \ln \lambda_n^{(s)}} = x_n^{(s)} \tag{47}$$

which is the mass balance relation from $\frac{\lambda_n^{(s)}}{z_\alpha} \frac{\partial z}{\partial \lambda_n^{(s)}} = x_n^{(s)}$ and we have the results: $x_n^{(s)} = \sum_{ijkl} \rho_{ijkl}$. These equations uniquely determine the parameters $\lambda_n^{(s)}$ based on Yang's proof.

After this transform, we can observe the key relation:

$$\frac{\partial \Psi_\alpha}{\partial T} = \frac{\partial \ln Z}{\partial T} = \sum_{i,j,k,l} \frac{(\prod_{s=1}^{S} \xi_s(ijkl)) \exp\left(-\frac{\varepsilon_{ijkl}}{k_B T}\right)}{z_\alpha} \cdot \frac{\varepsilon_{ijkl}}{k_B T^2} = \frac{1}{k_B T^2} \sum_{i,j,k,l} \rho_{ijkl} \varepsilon_{ijkl} = \frac{\varepsilon_\alpha}{k_B T^2} \tag{48}$$

Let's come back to free energy. Based on the previous calculation, we can overcome the integral with the calculated derivative. As the energy and the integral have additivity, we can directly separately expand and sum up the contribution from different types of clusters and perform the Legendre transform separately to get the final formula.

$$F(x,T) = -\left(k_B T \ln Z(x,\infty) + k_B T \int_\infty^T \frac{E}{k_B T^2} dT\right)$$
$$= -\left\{k_B T \ln Z(x,\infty) + kT \sum_\alpha a_\alpha m_\alpha [\Psi_\alpha(x,T) - \Psi_\alpha(x,\infty)]\right\} \tag{49}$$

$$\Psi_\alpha(x,T) = \ln z_\alpha - \sum_{s,n} x_n^{(s)} \ln \lambda_n^{(s)} \tag{50}$$

Here $\Psi(x,\infty)$ is the high temperature limit for the cluster formalism of the integral. We also notice $k_B T \ln Z(x,\infty)$ represents the high temperature limit as the Bragg-Williams approximation, then we can finally get the formula by just inserting all of them and this would give a same formula discussed before. As the high temperature limit for the most general case involves more technical details, we provide rigorous mathematical proof in the next section.

**S1.2 High-temperature limit**

We prove the most generalized form for the high temperature limit $\Psi_\alpha(x,\infty)$, with both multi-component case and variant clusters. This proof is important to supplement Yang's original proof [1] though we have already developed a new formalism to reconstruct it by grand canonical ensemble. It's especially significant for the multicomponent case, as Oates revived CSA but never provided rigorous proof for the multicomponent version of CSA [8,9] while Yang and Li focused on the binary case within their previous work and never derived this for the multicomponent case [1,2,10,11].

We need to calculate

$$\Psi(x,\infty) = \ln Z - \sum_{s,n} x_n^{(s)} \ln \lambda_n^{(s)} \tag{51}$$

$x_n^{(s)}$ and $\lambda_n^{(s)}$ should approach constant values as $T \to \infty$ at high temperature limit. Note here the derivation, $n$ is from 1 to $N-1$ to let the numbers of $x_n^{(s)}$ and $\lambda_n^{(s)}$ have the same number of indices. With the help of multinomial theorem, and the assumption under the high temperature limit, we have the partition function:

$$Z(T \to \infty) = \prod_s \left(1 + \sum_n \lambda_n^{(s)}\right) \tag{52}$$

As $\exp\left(-\frac{\varepsilon}{kT}\right) \to 1$ with the high temperature limit.

And the relation between $x$ and $\lambda$

$$\frac{\lambda_n^{(s)}}{\psi} \frac{\partial Z}{\partial \lambda_n^{(s)}} = x_n^{(s)} \tag{53}$$

We consider how this equation changes under the high temperature limit. For LHS, we insert the partition function at the high temperature:

$$\frac{\lambda_{n_0}^{(s_0)}}{Z} \frac{\partial Z}{\partial \lambda_{n_0}^{(s_0)}} = \frac{\lambda_{n_0}^{(s_0)} \prod_{s \neq s_0}\left(1 + \sum_n \lambda_n^{(s)}\right)}{\prod_s \left(1 + \sum_n \lambda_n^{(s)}\right)} \tag{54}$$

Then we have:

$$\frac{\lambda_{n_0}^{(s_0)} \prod_{s \neq s_0}\left(1 + \sum_n \lambda_n^{(s)}\right)}{\prod_s \left(1 + \sum_n \lambda_n^{(s)}\right)} = x_{n_0}^{(s_0)} \tag{55}$$

Simplify it we have:

$$\frac{\lambda_{n_0}^{(s_0)}}{1 + \sum_n \lambda_{n_0}^{(s)}} = x_{n_0}^{(s_0)} \tag{56}$$

For every possible $s_0, n_0$, we all have this equation. Then we add all the equations respect to the fixed $s_0$, we have:

$$\frac{\sum_n \lambda_n^{(s_0)}}{1 + \sum_n \lambda_n^{(s_0)}} = \sum_n x_n^{(s_0)} = 1 - x_N^{(s_0)} = A_{s_0} \tag{57}$$

We use $A$ to represent this sum.

Then we have:

$$\sum_n \lambda_n^{(s_0)} = A_{s_0}\left(1 + \sum_s \lambda_n^{(s_0)}\right) \tag{58}$$

$$\sum_n \lambda_n^{(s_0)} = \frac{A_{s_0}}{1 - A_{s_0}} \tag{59}$$

Insert this back to each equation,

$$\frac{\lambda_n^{(s_0)}}{1 + \sum_n \lambda_n^{(s_0)}} = x_{n_0}^{(s_0)} \tag{60}$$

Then we have:

$$\frac{\lambda_{n_0}^{(s_0)}}{1 + \frac{A_{s_0}}{1 - A_{s_0}}} = x_{n_0}^{(s_0)} \tag{61}$$

$$\lambda_{n_0}^{(s_0)} = \frac{x_{n_0}^{(s_0)}}{1 - A_{s_0}} \tag{62}$$

And

$$Z = \prod_s \left(1 + \sum_n \lambda_n^{(s)}\right) = \prod_s \frac{1}{(1 - A_s)} \tag{63}$$

Which are both under the high temperature limit $T \to \infty$

$$\Psi(x, \infty) = \ln Z - \sum_{s,n} x_n^{(s)} \ln \lambda_n^{(s)}$$

$$= -\sum_s \ln(1 - A_s) - \sum_{s,n} x_n^{(s)} \ln \frac{x_n^{(s)}}{1 - A_s}$$

$$= -\left(\sum_{s,n} x_n^{(s)} \ln x_n^{(s)} + \sum_s (1 - A_s) \ln(1 - A_s)\right) \tag{64}$$

If we allow $n$ comes from 1 to $N$ for $N$ component system, then we can see

$$\Psi(x, \infty) = -\sum_{s,n} x_n^{(s)} \ln x_n^{(s)} \tag{65}$$

with all the sites involved in such a cluster. This mathematical form is exactly what was postulated based on the binary's situation derived by Yang and Li [1,2] and is equivalent to what we derived in main text.

## S2. The relationship between site variables in FYL-CVM and correlation functions

We attempt to relate the site variables $\lambda$ from FYL-CVM with correlation functions in cluster expansion. Let's limit our discussion on the tetrahedron approximation for fcc binary system. For FYL-CVM, we have

$$\rho_{ijkl} = \frac{1}{Z_\alpha} \lambda^{(1)} \lambda^{(2)} \lambda^{(3)} \lambda^{(4)} \exp\left(-\frac{\varepsilon_{ijkl}}{kT}\right) \tag{66}$$

For the correlation function, they can be connected to the cluster probability as well. We want to derive the relation between $\lambda_i$ and $\xi_i$.

The general relation between basic cluster probability and the correlation function can be expressed as:

$$\rho_c(l) = \frac{1}{2^n}\left\{1 + \sum_{l'} V_{\{c\}}(l,l') \cdot \xi_{l'}\right\} \tag{67}$$

where $c$ represent different cluster configurations, $l$ indicates the cluster formed by n-lattice points while $l'$ is a subcluster contained in a cluster $l$, $V$ is the V-matrix [12]. This relation connects cluster probability and correlation functions linearly and generally. Thus, we have

$$\rho_{ijkl} = \frac{1}{2^4}\left\{\begin{array}{c} 1 + (i+j+k+l)\xi_1 + (ij+ik+il+jk+jl+kl)\xi_2 \\ +(ijk+ijl+ikl+jkl)\xi_3 + ijkl\xi_4 \end{array}\right\} \tag{68}$$

However, this is for the disordered phase. For the ordered phase, we have

$$\rho_{ijkl} = \frac{1}{2^4}\left\{\begin{array}{c} 1 + \left(i\xi_1^i + j\xi_1^j + k\xi_1^k + l\xi_1^l\right) \\ +\left(ij\xi_2^{ij} + ik\xi_2^{ik} + il\xi_2^{il} + jk\xi_2^{jk} + jl\xi_2^{jl} + kl\xi_2^{kl}\right) \\ +\left(ijk\xi_3^{ijk} + ijl\xi_3^{ijl} + ikl\xi_3^{ikl} + jkl\xi_3^{jkl}\right) + ijkl\xi_4 \end{array}\right\} \tag{69}$$

There are 16 equations to determine $\xi$ from $\lambda$. Assume we have coefficients matrix A for $\xi$, then we have

$$A\xi + \frac{1}{16} = b \tag{70}$$

where

$$b = \rho_{ijkl} = \frac{1}{z_\alpha}\lambda^{(1)}\lambda^{(2)}\lambda^{(3)}\lambda^{(4)}\exp\left(-\frac{\varepsilon_{ijkl}}{kT}\right) \tag{71}$$

Solve these linear equations with 15 variables and we get the $\xi(\lambda)$. In summary, there is no explicit relationship between $\xi$ and $\lambda$.

# Supplementary References